\documentstyle[12pt]{article}
\newcommand{\be}{\begin{equation}}
\newcommand{\ee}{\end{equation}}
\newcommand{\bea}{\begin{eqnarray}}
\newcommand{\eea}{\end{eqnarray}}

\begin{document} 

\baselineskip=18pt
\begin{center}
{\large\bf Collective Excitation of a Hybrid Atomic/Molecular Bose
Einstein Condensate$^{\ast}$ }
\end{center}
\begin{center}
Chi-Yong. Lin$^{\dag}$, M.S. Hussein$^{\ddag}$, A. F. R. de Toledo Piza
\end{center}
\begin{center}
{\it
Instituto de F\'{\i}sica, Universidade de S\~ao Paulo, CP 66318\\  
CEP 05389-970, S\~ao Paulo, \ SP, \ Brazil}
\end{center}

\begin{center}
E. Timmermans
\end{center}
\begin{center}
{\it
T-4, Los Alamos National Laboratory\\
Los Alamos, NM 87545, \ U.S.A.}
\end{center}

\vskip .3cm
\baselineskip=18pt
\begin{center}
{\bf ABSTRACT}
\end{center}
\vskip 0.1cm
\indent{The} collective excitation of hybrid atomic-molecular
condensate are studied using variational method. 
The dipole response of the system is studied in detail. 
We found that the out-of-phase dipole oscillation frequency increasing
slowly with the detuning parameter and is only few times large than
the trap frequency making the experimental observation of this mode of
vibration feasible.
\vskip 1.5cm

\vfill
\hspace{\fill}

\noindent\makebox[66mm]{\hrulefill}

\footnotesize 
$^{\ast}$
Supported in part by FAPESP, Brazil.

$^{\dag}$
Supported by FAPESP, Brazil.

$^{\ddag}$
Supported in part by CNPq, Brazil.
\newpage
\normalsize
\baselineskip=30pt
\indent
The recent realization of Bose-Einstein condensation (BEC) in a
dilute gas of alkali atoms has open a new opportunity for the
theoretical and experimental investigation in quantum degenerate
many-body systems \cite{KDS99,DGPS99}.
In contrast to superfluid helium, these weakly interacting gases are much
more amenable to the theoretical prediction and quantative analysis.
After the observations of such condensates, a new generation of more
complex experiments has been done involving the production of multiple
species BEC.  
Such mixture may consist of different hypefine states of same atom
or different spin state of same stomic species.
In the former case the composition of mixture remains fixed and in the
later case the internal conversion leads to important new phenomena.  
It has been proposed that the interation that bring a binary atom
system to an intermediate state molecule in the Feshbach resonance,
recently observed in BEC\cite{In98,St99}, create in the dilute atomic
Bose-Eintein condensate (BEC) a second molecular condensate
component\cite{TTHK98}. The communication between the atomic and molecular
condensates ensues through a Josephson-like oscillation of atom
pair. It was found that the ground state of this hybrid condensate is
a dilute BEC with liquid-like properties of a self-determined
density. These findings were obtained using a homogeneous gas
approximation (no confining potential). 
In this work we investigate the collective excitations of the hybrid
system of atoms and  molecules subjected to an appropriate trap. 
In particular we derive an appproximate expression for the dipole
frequency of the out-of-phase oscillations of the atomic and molecular
condensate centers of mass and investigate its dependence on the
detuning and Feshbach resonance parameters.

\indent
The mean-field hamiltonian density describing the trapped system is
\bea
{\cal H}&=&\psi_a^*\left[-\frac{\hbar^2\nabla^2}{2m}
     +V_a({\bf r})+\frac{\lambda_a}{2}\psi_a^*\psi_a\right]\psi_a
       +\psi_m^*\left[-\frac{\hbar^2\nabla^2}{4m}
     +V_m({\bf r})+\varepsilon+\frac{\lambda_m}{2}\psi_m^*\psi_m\right]\psi_m \nonumber\\
&+&\lambda\psi_a^*\psi_a\psi_m^*\psi_m
+\frac{\alpha}{\sqrt{2}}
 \Bigl[\psi^*_m\psi_a\psi_a+\psi_m\psi^*_a\psi^*_a\Bigr] \;,
\eea
where $\psi_{a(m)}$ denote the macroscopic wavefunction of the atom
(molecule) condensate, $\varepsilon$ is 
the detuning parameter, $\alpha$ is the Feshbach resonance parameter
introduced in Ref.\cite{TTHK98}, 
and we adopt here a simple form for
the atomic (molecular) confining potentials, 
$V_a=\frac{\omega_{a}^2}{2m}(\lambda^{a}_x
x^2+\lambda^{a}_{y}y^2+\lambda^{a}_{z}z^2)$,
$V_m=\frac{\omega_{m}^2}{2(2m)}(\lambda^{m}_x
x^2+\lambda^{m}_{y}y^2+\lambda^{m}_{z}z^2)$.
As usual, the Gross-Pitaevskii equations of motion for this model can
be derived from the action  
$\Gamma=\int dt \;
d^3r\frac{i}{2}\Bigl[\psi_a^*\dot\psi_a-\dot\psi_a^*\psi_a+\psi_m^*\dot\psi_m-\dot\psi_a^*\psi_a\Big]
-\int dt\; E [\psi_a,\psi_m ] \;, 
$
\noindent where $E[\psi_a,\psi_m]=\int d ^3r \; {\cal H}\;$.

\indent
In order to investigate the general properties of the hybrid system described by
${\cal H}$, Eq.(1), we employ the time-dependent variational Gaussian
ansatz \cite{BP96} 
\bea
\psi_a(r,t)&=&\frac{N_a^{1/2}}{\pi^{3/4}q^{3/2}}
\prod_{j=x,\;y,\;z}
e^{ -\bigl[\frac{1}{2q^2_j(t)} -i\frac{p_j(t)}{q_j(t)}\bigr]
\bigl[r_j-r^{c}_{j}(t)\bigr]^2
+i\pi^{c}_{j}(t)\bigl[r_j-r^{c}_{j}(t)\bigr]+i\theta_a(t)}\\
\psi_m(r,t)&=&\frac{N_m^{1/2}}{\pi^{3/4}Q^{3/2}}
\prod_{j=x,\;y,\;z}
e^{ -\bigl[\frac{1}{2Q^2_j(t)} -i\frac{P_j(t)}{Q_j(t)}\bigr]
\bigl[r_j-R^{c}_{j}(t)\bigr]^2
+i\Pi^{c}_{j}(t)\bigl[r_j-R^{c}_{j}(t)\bigr]+i\theta_m(t) }
\eea
where 
$q_j,p_j,r^{c}_{j},\pi^{c}_{j}$ ($Q_j,P_j,R^{c}_{j},\Pi^{c}_{j}$) are
our variational parameters for the atomic (molecular) condensate
wavefunctions and we have introduced the notation
$q=(q_xq_yq_z)^{1/3}$ and $Q=(Q_xQ_yQ_z)^{1/3}$. 
Thus, $\psi_a(\psi_m)$ is a Gaussian 
centered at $r^{c}_{j}$ $(R^{c}_{j})$ with a width $q_j$ $(Q_j)$ and conjugate
momenta $\pi^{c}_{j}$ $(\Pi^{c}_{j})$ and $p_j$ $(P_j)$ respectively. 
In addition, we shall use as variational parameters the numbers of the  atoms, 
$\int d^3r|\psi^*_{a}|^2 = N_{a}$, and  molecules, 
$\int d^3r|\psi^*_{m}|^2 = N_{m}$, subject to the
constraint $N_a+2N_m=N$ (fixed).

\indent{The} action in this approximaion is therefore
\be
\Gamma \;=\;\int dt \; \Bigl\{ \;
\theta_a\dot N_a\;+\; \theta_m\dot N_m 
+\sum_{j=x,\;y,\;z} N_a \pi^c_j\dot r^c_j 
+\frac{N_a}{4}\bigl(p_j\dot q_j - \dot p_j q_j\bigr)
\Bigr. \nonumber\\
N_m \Pi^c_j\dot R^c_j
+\frac{N_m}{4}\bigl(P_{j}\dot Q_{j}-\dot P_{j}Q_{j}\bigr)-E\Bigr\}
\ee

\noindent
The total energy has the general form $E=E_a+E_m+E_i+E_{_{FR}}$,
where $E_a(E_m)$ are the usual Gaussian Gross-Pitaevskii energy for a
single atomic(molecular) condensates \cite{BP96},
$E_{a}=N_a\sum_{j=x,y,z}
\bigl[\frac{1}{2m}\pi^2_{j}+\frac{1}{4m}p^2_{j}+\frac{\hbar^2}{4m}q^{-2}_{j}
+\frac{m}{2}\omega_{a_j}^2(r^c_j)^2+\frac{m}{4}\omega_{a_j}^2q^2_{j}\bigr]
+\frac{\lambda_aN^2_a}{2(2\pi)^{3/2}}q^{-3}$
($\pi_j\rightarrow\Pi_j$ and etc for $E_m$). 
The other two terms are, respectively, the elastic and inelastic
collisional contribution between the condensates, 
$E_{i}=\frac{\lambda N_aN_m}{\pi^{3/2}}
\large\{\prod_{j=x,\;y,\;z}exp[-\frac{(r^c_j-R^c_j)^2}{q^2_{j}+Q^2_{j}}]
(q^2_{j}+Q^2_{j})^{-1/2}\large\}$, 
$E_{_{FR}}=\varepsilon N_m +
\frac{\alpha N_aN^{1/2}_m}{2^{1/2}\pi^{3/4}q^3Q^3}
\large\{e^{2i\theta}
\prod_{j=x,\;y,\;z}exp[\frac{b_j^2}{4a_{j}}-c_j]a_j^{-1/2}
+{\rm c.c.}\large\}$,
where we have used the notations:
$a_j=B^*_j+2\beta_j$, $b_j=-2B^*_jD_j-4\beta_j r^c_j+i\Pi_j-2\pi_j$,
$c_j=B^*_jD^2_j+2\beta_j (r^c_j)^2-i\Pi_jR^c_j-2\pi_jr^c_j$,
being $a_j=B^*+2\beta$, $2B=Q^-2_j-P_jQ^{-1}$,
$2\beta=q^-2_j-p_jq_j^{-1}$. 
The variational equations are then obtained from $\delta\Gamma=0$
leading a set of first order differential equation of the form 
\be
\sum_{k}{\bf S}_{ik}\; \dot X_k = \; \frac{\partial E}{\partial X_i}\;,
\ee 
where ${\bf S}$ is an appropriate matrix.
These classical equations of motion give a simple picture for
the time evolution of the atomic (molecular) condensate wavefunction
parametrized by its widths and centers of mass and the dynamics of
inter-condensate tunneling driven by the FR interaction.
The ground state energy, $E=\int d ^3r \; {\cal H}\;$, on the other
hand is determined by the solution of Eq.(5) with $\dot X_k=0$. 
The small amplitude oscillation dynamics is described by an expansion
around the equilibrium solution. Thus, if $X^{(0)}$ is the static equilibrium
solution and $X^{(1)}$ is the perturbation, we have 
\be
i\omega\;{\bf K}\; X^{(1)} = {\bf A} \; X^{(1)}\;,
\ee
where
${\bf A}_{ij}=\left. \frac{\partial^2 E}{\partial X_{i}\partial X_{j}}\right|_{(0)}$
and ${\bf K}$ is an appropriate $26\times26$ inertia matrix. 
The set of 26
coupled equations is diagonalized for the frequencies, which
come out in pair of opposite parities. 
The thirteen physical frequencies can be grouped into three
sets: one corresponds to the motion of the centers of mass of
atomic/molecular gases; one defines the spatial
frontiers of the atomic/molecular condensate and one describes
the Josephson tunnelling effects between the two condensates. 
The frequencies associated with the Josephson tunneling were found in
Ref.\cite{TTHK98} (no trap) to be 
$\omega(X_a=\frac{N_a}{N},n)=\pm
n\lambda[4X_a+Xa\sqrt{1-X_a}+\frac{X^2_a}{1-X_a}]$.

\indent{An} important new phenomenon in the hybrid
condensate physics appears when the system is subjected to the
confining trap namely the in-phase and out-of-phase dipole
oscillations of the two centers of mass, what results four excitation
modes when cylindrically symmetric trap is used. 
The dipole oscillation of the center of mass of an atomic BEC is entirely
determined by the frequency of the trap and does not depend on the
interatomic interaction \cite{DGPS99}.
If, on the other hand, we allow atoms and
molecules, with their own confining traps and with Josephson type
oscillations, then presumably the dipole mode of oscillation of the
respective center of masses will depend on the Feshbach resonance
interaction parameter, $\alpha$, and on the detuning parameter
$\varepsilon$. 
If so, then there is an interesting phenomenon to be explored
experimentally. 
An analogous situation is discussed in a homogeneous hybrid
condensate (no trap). In that case, the low k-modes are, as experted for
superfluids, longitudinal waves, i.e. the oscillating particle current
is in the direction of propagation ({\bf k}). One mode, in which both the atomic
and molecular particle current move in the same direction, i.e. a
Goldstone mode and its excitation energy vanishes as ${\rm
k}\rightarrow0$. The other mode has atomic and 
molecular currents moving in opposite direction like a plasmon mode
and it has a gap: it goes to a finite value which depends on the
interaction parameters as ${\rm k}\rightarrow0$. 
In our case we would expect two similar modes: one gives an energy
$\omega_{trap}$ if atoms and molecules experience the same trap
potential as in the case of the single condensate and another mode
that has higher energy, which depends mostly on the interaction
parameters. In the following we quantify these observations using our
model. 

\indent{For} simplicity we consider first the case of isotropic trap,
$\lambda_x=\lambda_y=\lambda_z=1$, for both atoms and molecules.
In this case, the following trivial {\underline equilibrium} solution
is found
\be
r_c=\pi_c=R_c=\Pi_c=0\;.
\ee

\noindent{In} addition, it is easy to show that 
$\frac{\partial^2 E}{\partial X_{c}\partial X_{w}}=0$ 
where $X^c$ are generic coordinates of the subspaces corresponding the
centers of mass and  $X^{w}$ represent the other degrees of freedom of
the variational space. 
Thus, the matrix $A$ reduces to the form
{\tiny $ \left(\begin{array}{cc}C&0\\0&W\end{array}\right) $}, where
$C_{ik}=\frac{\partial^2 E}{\partial X^{c}_i \partial X^{c}_k}$
and
$W_{ik}=\frac{\partial^2 E}{\partial X^{w}_i \partial X^{w}_k}$.
\normalsize
\baselineskip=30pt
As a consequence,  $X_c$ decouple from $X_w$ at $X^{0}$
and the dynamics of the centers of mass system is described by 
two coupled oscillator equations. 
Eliminating $\pi^{(0)}_c$ and $\Pi^{(0)}_c$ we find  
\bea
N_a\;\ddot{r}_{_c}&=&-[N_a\;\omega_{a}^2+\;\Delta\omega^2]\;r_{_c}
+\Delta\omega^2\;R_{_c}\\
2N_m\;\ddot{R}_{_c}&=&-[2N_m\;\omega_{a}^2+\Delta\omega^2]\;R_{_c}
+\Delta\omega^2\;r_{_c}
\eea

\noindent{where} the frequency shift $\Delta\omega$ is produced by
the interaction between the atomic and molecular condensates and thus
it contains the $\alpha-$ and $\varepsilon-$ dependence. 
This is related to the curvature of $E^{i}$ at $X_0$ by the following
simple expression 
\be
\Delta\omega^2\;=\;m\omega_a^2
\Bigl(\frac{\partial^2 E^{i}}{\partial\pi^2_{c}}\Bigr)
+\frac{1}{m}\Bigl(\frac{\partial^2 E^{i}}{\partial r^2_{c}}\Bigr)
+\Bigl(\frac{N}{N_a}+\frac{N}{2N_m}\Bigr)
 \Bigl(\frac{\partial^2 E^{i}}{\partial\pi^2_{c}}\Bigr)
 \Bigl(\frac{\partial^2 E^{i}}{\partial r^2_{c}}\Bigr)
\ee
\noindent 
The two coupled equations (8) and (9) can be uncoupled in favor of the
normal modes, $\eta\equiv N_a r_{c}+2N_m R_{c}$ and 
$\xi\equiv r_{c}-R_{c}$, whose frequencies  are, respectively, 
\be
\Omega^2_{-}=\omega^2_a
\;,\hspace{1.0cm}
\Omega^2_{+}=\omega^2_a+\Delta\omega^2\frac{N}{2N_m(N-2N_m)}
\ee
The mode $\Omega_{-}$ is the in-phase solution where the centers of
mass of atomic and molecular gases move in the same direction and
is independent of the interactions \cite{DGPS99}. 
On the other hand, the out-of-phase dipole oscillation solution
$\Omega_{+}$, depends strongly on the composition of the system, which
is self-determined by the equilibrium condition $X_0$. 
From the previous discussion we have
$\frac{\partial^2 E^{i}}{\partial\pi^2_{c}} \sim N_aN_m$ and
$\frac{\partial^2 E^{_{_{FR}}}}{\partial\pi^2_{c}} \sim N_aN_m^{1/2}$. 
Thus, when $N_m<<N$, one has $\Omega_{+} \rightarrow N_m^{-1/2}$. 
This result is illustrated in Fig.1, where $\Omega_{+}$ increases
dramatically for solutions with small number of molecules.

\indent
We turn next to the quadrupole response of the system. In Fig.2 we
show the three corresponding frequencies    
The lowest frequency describes the oscillations $q$, $Q$ and $N_a$ in
the same direction. 
The other two modes describe the out-of-phase solution of $q$ against
$Q$, which leads to migration of atomic condensate to molecular
condensate (dash-dotted plot) and vice-verse (dotted curves ). 
This can also be represented by the current $\vec j_a={\bf\nabla}\rho_a$
($\vec j_m={\bf\nabla}\rho_m$) obeying continuity equation 
${\bf\nabla}\cdot (j_a+2j_m)=0$.
These modes, again, are strongly dependent on the particular
equilibrium point of the system. 

\indent{In} conclusion, we have discussed in this paper a variational
calculation of the small amplitude dynamics of a hybridatom/molecule
Bose Einstein condensate.
The out of phase dipole mode of oscillation of center of mass of the
two condensed gases is considered in detail. 
For reasonable values of the physical parameters (the Feshbach
resonance strength parameter, densities of gases, etc), we have found
that the dipole frequency depends approximatelly quadratically on the
detuning parametes, $\varepsilon$, at small value of the latter
attaining an almost linear dependence for higher values of
$\varepsilon$. 
This frequency is less than an order of magnitude large than the trap
frequencyfor values of $\varepsilon$ of several KHz.

\newpage
\begin{center}
{\bf Figure Captions}
\end{center}

\noindent{Figure} 1. 
The out-of-phase dipole frequency as function of detuning parameter,
$\varepsilon$, for the hybrid condensate in unit of trap frequency. 
The parameters are: 
$\lambda_a=\lambda_m=\lambda=100 m\mu^3 s^-1$, 
$N=10^4$, $V_a=2\pi*50Hz$, $\alpha=200m\mu^{3/2}s$.
See text for details.

\noindent{Figure} 2. 
The three eigenfrequencies for the $q$, $Q$ and $N_a$ degrees of
freedom. 
Same parameters are used as in Fig.1. 
Lower curve, $\omega(q+Q+N_a)$; middle curve, $\omega(q-Q+N_a)$; upper
curve, $\omega(q-Q-N_a)$. See text for details.

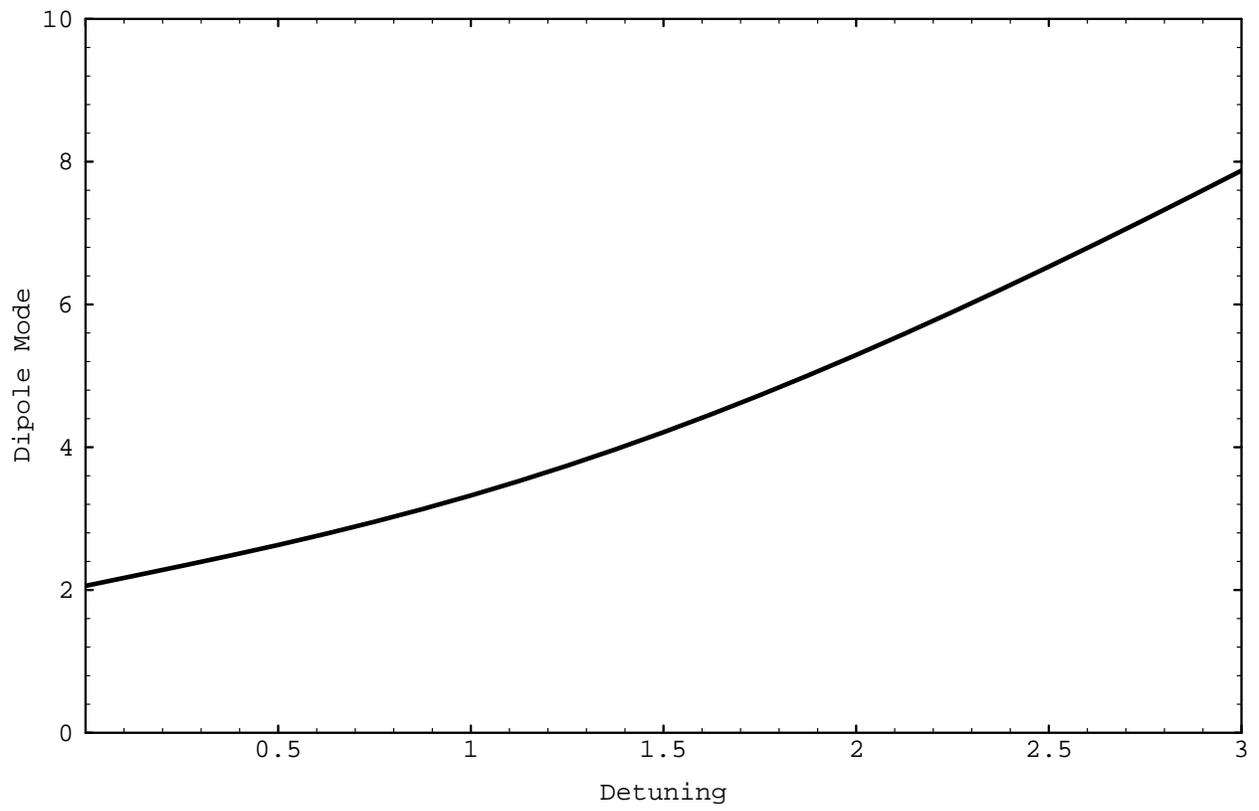
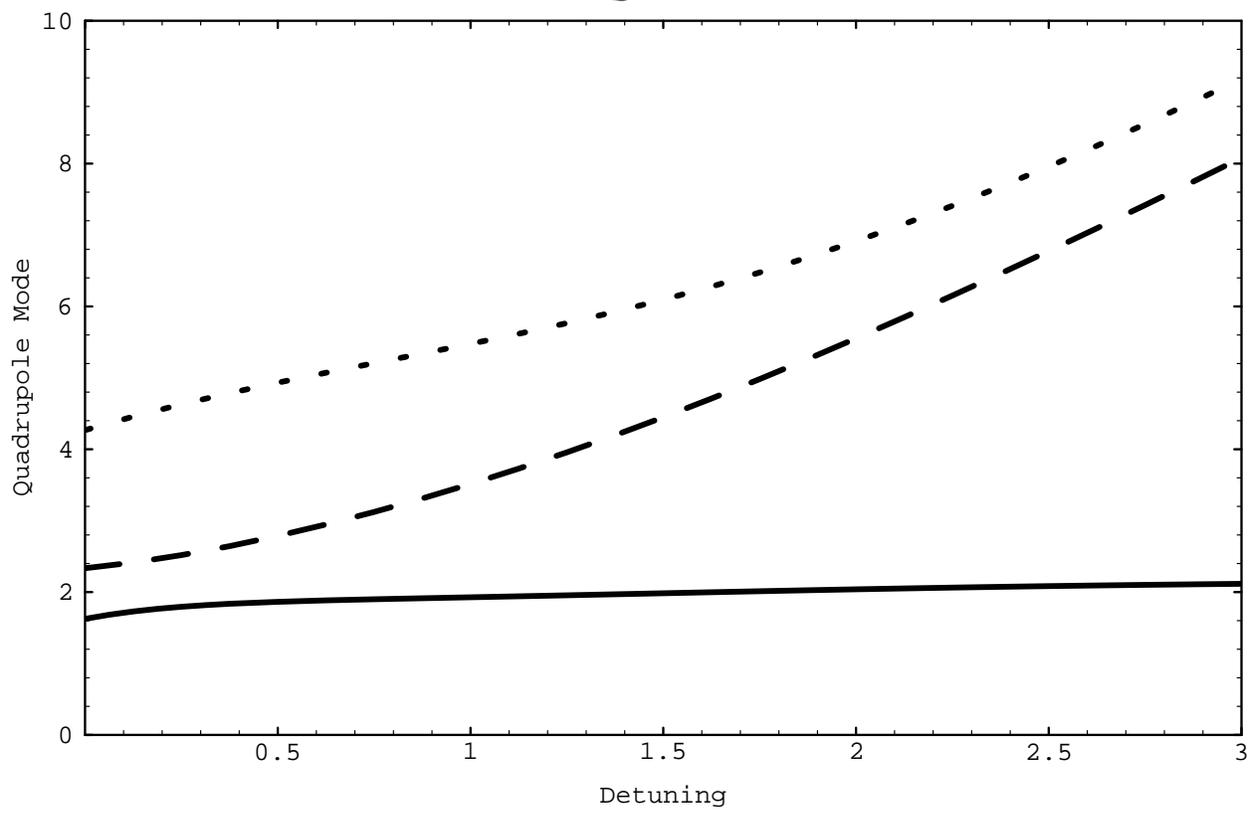
\end{document}